\documentclass[useAMS,usenatbib]{mnras}
\usepackage{multirow}
\usepackage{epsfig}
\usepackage{subfigure}
\usepackage{graphicx}
\usepackage{amsmath}
\usepackage{color}

\makeatletter
\let\jnl@style=\rm
\def\ref@jnl#1{{\jnl@style#1}}

\def\aj{\ref@jnl{AJ}}                   
\def\araa{\ref@jnl{ARA\&A}}             
\def\apj{\ref@jnl{ApJ}}                 
\def\apjl{\ref@jnl{ApJ}}                
\def\apjs{\ref@jnl{ApJS}}               
\def\ao{\ref@jnl{Appl.~Opt.}}           
\def\apss{\ref@jnl{Ap\&SS}}             
\def\aap{\ref@jnl{A\&A}}                
\def\aapr{\ref@jnl{A\&A~Rev.}}          
\def\aaps{\ref@jnl{A\&AS}}              
\def\azh{\ref@jnl{AZh}}                 
\def\baas{\ref@jnl{BAAS}}               
\def\jrasc{\ref@jnl{JRASC}}             
\def\memras{\ref@jnl{MmRAS}}            
\def\mnras{\ref@jnl{MNRAS}}             
\def\pra{\ref@jnl{Phys.~Rev.~A}}        
\def\prb{\ref@jnl{Phys.~Rev.~B}}        
\def\prc{\ref@jnl{Phys.~Rev.~C}}        
\def\prd{\ref@jnl{Phys.~Rev.~D}}        
\def\pre{\ref@jnl{Phys.~Rev.~E}}        
\def\prl{\ref@jnl{Phys.~Rev.~Lett.}}    
\def\pasp{\ref@jnl{PASP}}               
\def\pasj{\ref@jnl{PASJ}}               
\def\qjras{\ref@jnl{QJRAS}}             
\def\skytel{\ref@jnl{S\&T}}             
\def\solphys{\ref@jnl{Sol.~Phys.}}      
\def\sovast{\ref@jnl{Soviet~Ast.}}      
\def\ssr{\ref@jnl{Space~Sci.~Rev.}}     
\def\zap{\ref@jnl{ZAp}}                 
\def\nat{\ref@jnl{Nature}}              
\def\iaucirc{\ref@jnl{IAU~Circ.}}       
\def\aplett{\ref@jnl{Astrophys.~Lett.}} 
\def\apspr{\ref@jnl{Astrophys.~Space~Phys.~Res.}}
\def\bain{\ref@jnl{Bull.~Astron.~Inst.~Netherlands}}
\def\fcp{\ref@jnl{Fund.~Cosmic~Phys.}}  
\def\gca{\ref@jnl{Geochim.~Cosmochim.~Acta}}   
\def\grl{\ref@jnl{Geophys.~Res.~Lett.}} 
\def\jcp{\ref@jnl{J.~Chem.~Phys.}}      
\def\jgr{\ref@jnl{J.~Geophys.~Res.}}    
\def\jqsrt{\ref@jnl{J.~Quant.~Spec.~Radiat.~Transf.}}
\def\memsai{\ref@jnl{Mem.~Soc.~Astron.~Italiana}}
\def\nphysa{\ref@jnl{Nucl.~Phys.~A}}   
\def\physrep{\ref@jnl{Phys.~Rep.}}   
\def\physscr{\ref@jnl{Phys.~Scr}}   
\def\planss{\ref@jnl{Planet.~Space~Sci.}}   
\def\procspie{\ref@jnl{Proc.~SPIE}}   

\makeatother

\newcommand{\xmm}{{\it XMM-Newton}}

\newcommand{\nustar}{\textit{NuSTAR}}

\def\Fevc{\ion{Fe}{xxv}}
\def\Fevs{\ion{Fe}{xxvi}}
\def\grsdn{GRS~1915+105}
\def\axj{AX~J1745.6-2901}

\title[Photoionization instability in AX~J1745.6-2901]{Photoionization instability of the Fe K absorbing plasma in the neutron star transient AX~J1745.6-2901}

\author[Stefano Bianchi, et al.]{Stefano Bianchi$^1$\thanks{E-mail: bianchi@fis.uniroma3.it (SB)}, Gabriele Ponti$^{2}$, Teo Mu\~noz-Darias$^{3,4}$, Pierre-Olivier Petrucci$^{5}$\\
$^1$Dipartimento di Matematica e Fisica, Universit\`a degli Studi Roma Tre, via della Vasca Navale 84, 00146 Roma, Italy\\
$^2$Max-Planck-Institut f{\"u}r Extraterrestrische Physik, Giessenbachstrasse, D-85748, Garching, Germany\\
$^{3}$Instituto de Astrof\'isica de Canarias (IAC), V\'ia L\'actea s/n, La Laguna 38205, S/C de Tenerife, Spain\\
$^{4}$Departamento de Astrof\'isica, Universidad de La Laguna, La Laguna, E-38205, S/C de Tenerife, Spain\\
$^{5}$Univ. Grenoble Alpes, CNRS, IPAG, F-38000 Grenoble, France\\
}

\begin{document}

\maketitle

\label{firstpage}

\begin{abstract}
\axj\ is a Low Mass X-ray Binary with an accreting neutron star, showing clear evidence for highly ionized absorption. Strong ionized Fe K$\alpha$ and K$\beta$ absorption lines are always observed during the soft state, whereas they disappear during the hard states. We computed photoionization stability curves for the hard and the soft state, under different assumptions on the adopted spectral energy distributions and the physical parameters of the plasma. We observe that the ionized absorber lies always on a stable branch of the photoionization stability curve during the soft state, while it becomes unstable during the hard state. This suggests that photoionization instability plays a key role in defining the observable properties of the ionized absorber. The same process might explain the disappearance of the high ionization absorber/wind, during the hard state in other accreting neutron stars and black holes. 
\end{abstract}

\begin{keywords}
X-rays: binaries -- X-rays: individual: \axj\ -- accretion, accretion discs -- stars: neutron -- black hole physics --
stars: winds, outflows
\end{keywords}

\section{Introduction}

Outflows, in the form of collimated jets and wide angle winds, are ubiquitous components of accretion onto compact objects \citep{Fender2004,Ponti2012b}. Indeed, a compact steady jet, best traced in the radio 
band, is a universal characteristic of the observed canonical hard state in accreting black holes (BHs) and neutron stars \citep[NSs;][]{Fender2004,Fender2016a}. As the source transits from the hard to the soft state, the jet becomes erratic, showing major radio flares that can be resolved into discrete ejecta moving at relativistic speed \citep{Mirabel1994,Hjellming1995,Fender1999}. On the other hand, jets are quenched in the soft state \citep{Fender1999,Fender2004,Fender2009,Corbel2000,Corbel2004,Tudose2009,Coriat2011,Russell2011}. 
Winds, traced by high ionization absorption lines such as \Fevc\ and \Fevs, are another fundamental component of accretion \citep[see][and references therein]{Ponti2016}. Interestingly, the evolution of winds seems to be complementary to the one of jets. Indeed, winds are a universal component during the soft state, while they are typically not observed during the canonical hard state \citep{Neilsen2009,Ponti2012b,Ponti2014}. 

The nature of the apparent mutual exclusion between the presence of the jet and the wind is still highly debated. It was initially suggested that jets and winds might be two different manifestations of the same outflow, which would be classified as a jet when collimated and as a wind otherwise \citep{Neilsen2009}. However, the simultaneous detection of a jet and a wind in a few accreting BHs and NSs appears to exclude this possibility \citep[e.g., V404~Cyg, one observation of \grsdn, GX~13+1, Sco~X-1 and Cir~X-1;][]{Munoz-Darias2016,Munoz-Darias2017,Homan2016}. 

It is possible that the observed phenomenology of the outflow is driven by the state of the accretion flow \citep{Ponti2012b,Ponti2016}. For example, it was proposed that the disappearance of the wind/ionized-absorption during the hard state might be due to over-ionization, as a consequence of the much harder illuminating spectral energy distribution (SED). However, it has been shown that in many instances (including the case of \axj) the ionized absorption does not vanish because of over-ionization \citep[e.g.][]{Miller2012,Neilsen2012,Ponti2015}. 

On the other hand, the different illuminating SEDs produced by the accretion flow over the various states could have more subtle effects, making the wind/absorption photoionization unstable \citep{Chakravorty2013a}. 
Indeed, it has been shown that a very hard X-ray SED, typically observed during the canonical hard state, can generate photoionization instabilities once it illuminates the wind, preventing it from lying within the range of 
parameter space typically observed during the soft state \citep{Chakravorty2013a}. We plan here to check whether this is indeed the case thanks to extensive \xmm\ and \nustar\ observations of one of the best monitored Low Mass X-ray Binary (LMXB), \axj, allowing us to detail the evolution of the SED.

\section{The SEDs of \axj}
\label{seds}

\axj\ is a transient accreting NS, classified as an atoll source \citep{Gladstone2007}. These systems accrete at mid to low accretion rates ($\la30\%$ the Eddington luminosity), showing very similar hard and soft accretion states (i.e. timing and spectral properties), as well as accretion/ejection coupling, to those observed in BH transients \citep{Munoz-Darias2014,Motta2017}. The orbital plane and accretion disc in this system are observed at a high inclination, showing dips, eclipses, and clear evidence for highly ionized absorption \citep{Hyodo2009,Ponti2015}. Ponti et al. (2017) confirmed, with a large number of X-ray observations, the ubiquitous presence of intense ionized Fe K$\alpha$ and K$\beta$ absorption lines during the soft state, which disappear with tight upper limits during the hard states. As explicitly shown by \citet{Ponti2015}, if the plasma retained its physical properties (i.e. density and location) in the hard states, it would remain easily detectable with the available observations.

\begin{figure}
\includegraphics[width=\columnwidth]{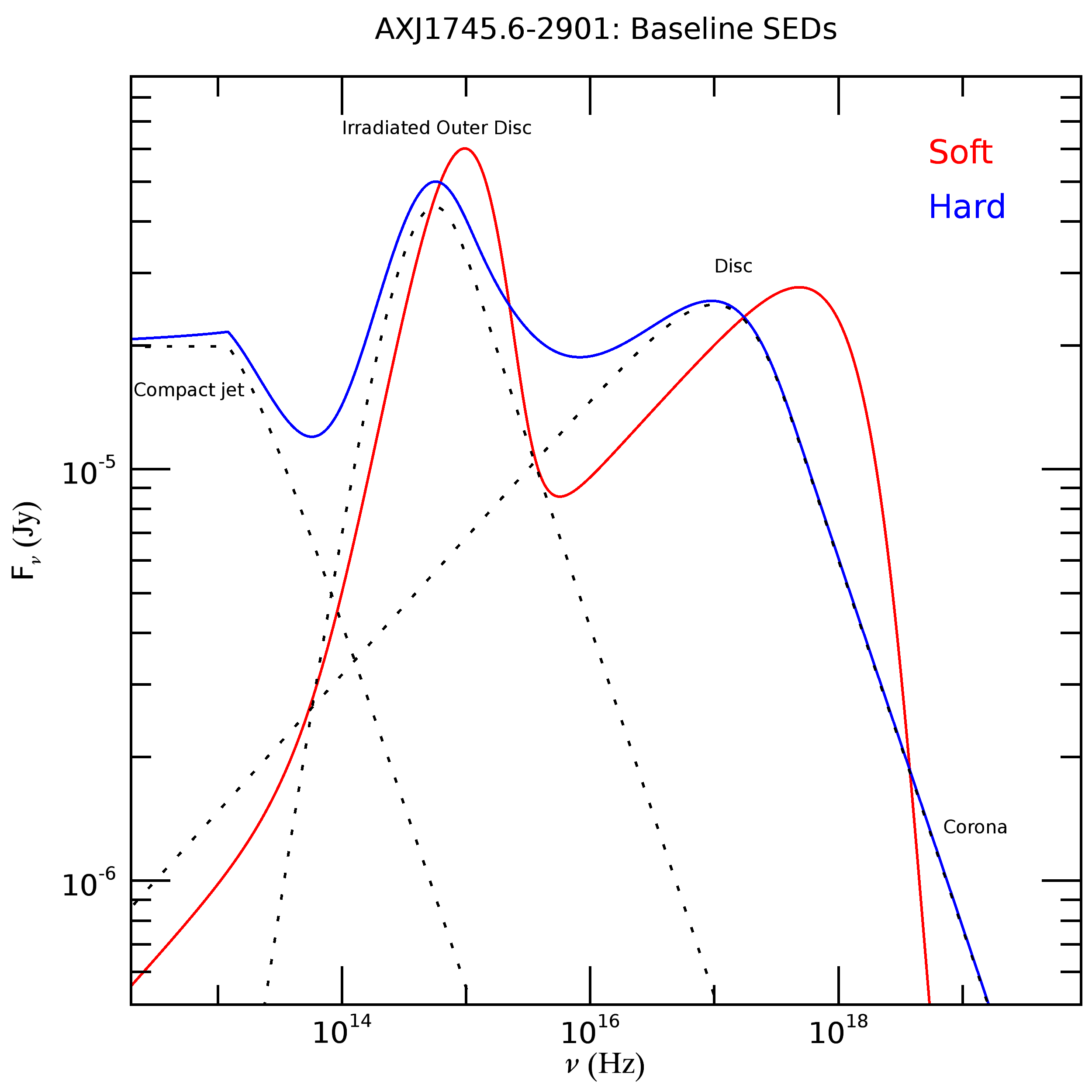}
\caption{The intrinsic \textit{baseline} SEDs of \axj\ during the soft (red line) and hard states (blue line). The key components are labelled. See text and \citet{Ponti2015} for details.}
\label{baselineseds}
\end{figure}

In order to compute the self-consistent photoionization model in their analysis, \citet{Ponti2015} constructed Spectral Energy Distributions (SEDs) of the central source illuminating the plasma, based on \nustar\ data, whose 3-78 keV energy band covers the most relevant part of the spectrum for very highly ionized plasma. In this energy band, the intrinsic SEDs can be modelled by a multi-colour disc emission and a power law arising from Comptonization of its seed photons. In the soft state, the inner temperature of the dominant disc component is well constrained ($kT=2$ keV), while only a very weak and steep power law component is observed. On the other hand, the hard state is dominated by the power law ($\Gamma=1.9$), and the disc emission is basically unconstrained, due to the very large Galactic absorption in the line of sight. An inner temperature of the disc $kT=0.3$ keV is assumed, together with a fraction $f_{sc}=0.8$ of Comptonized seed photons. For the optical and infrared bands of the SEDs, emission from the irradiated disc is modelled by a blackbody with temperature $T=15\,000$ and $7\,000$ K, for the soft and hard state, respectively. The contribution at radio-to-infrared frequencies from a compact jet is only added in the hard state. The resulting SEDs, shown in Fig.~\ref{baselineseds}, will be referred to as the \textit{baseline} SEDs in this paper. The ratio of the total ionizing luminosities of these SEDs is $L_h/L_s\sim0.155$.  

\begin{figure*}
\includegraphics[width=\columnwidth]{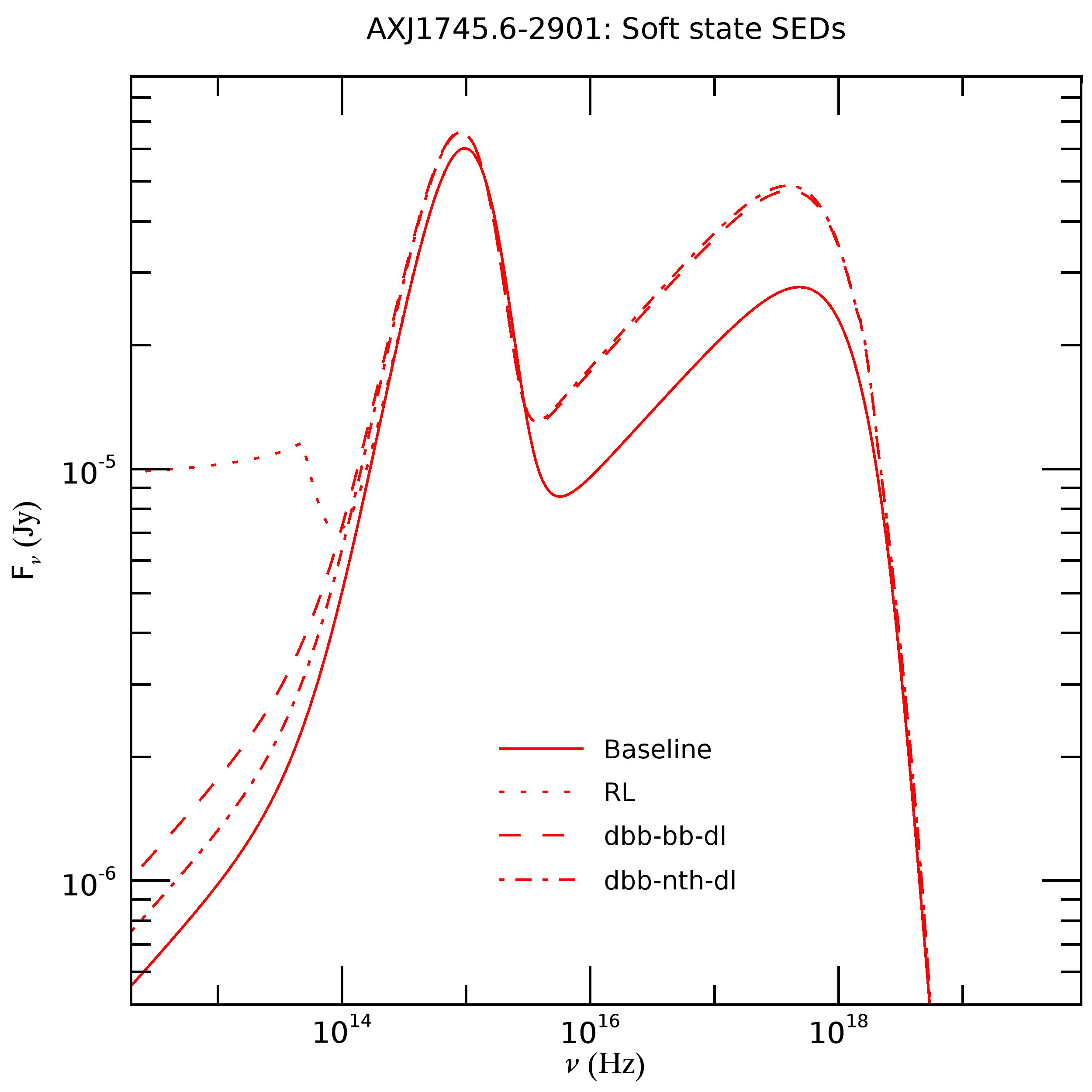}
\includegraphics[width=\columnwidth]{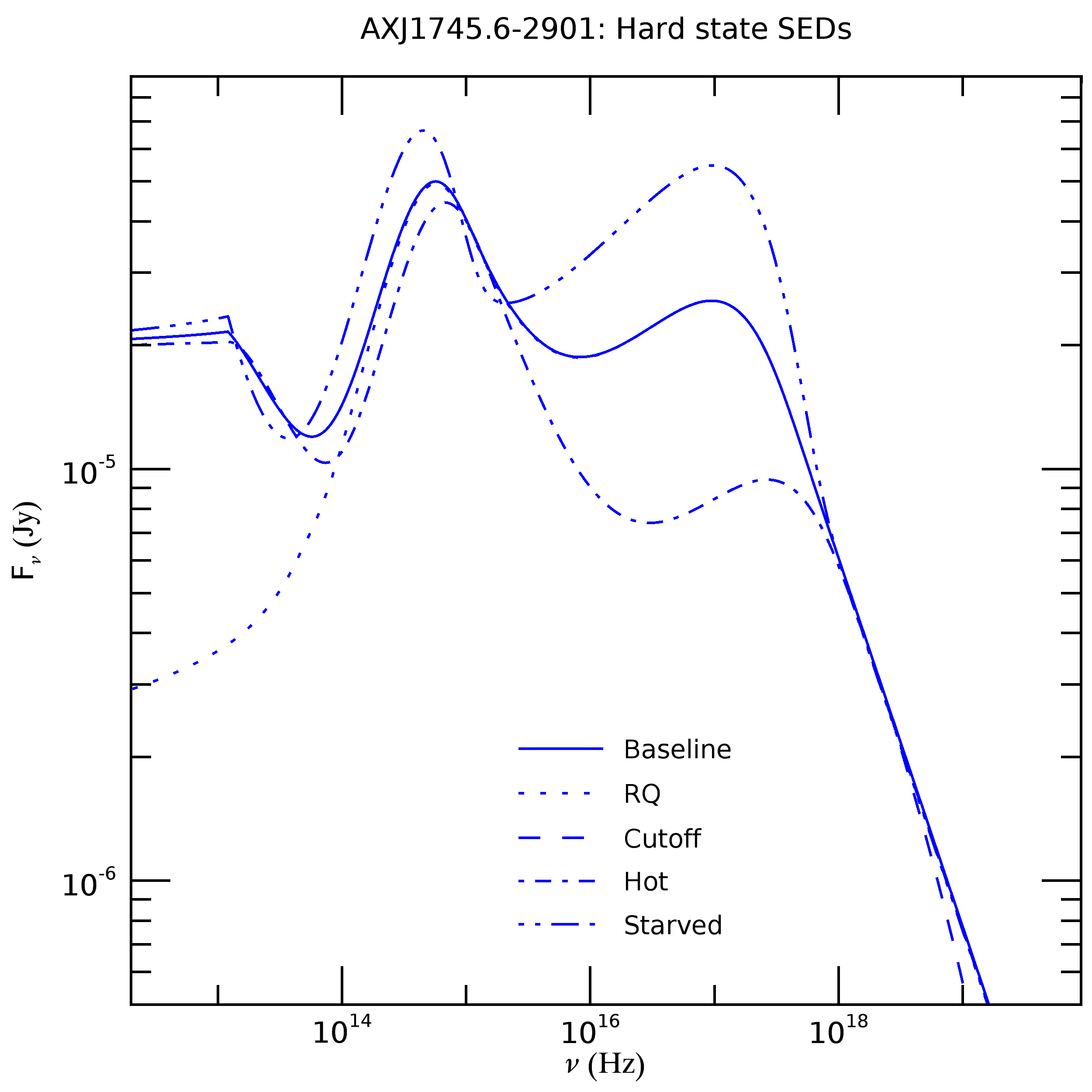}
\caption{Other intrinsic SEDs of \axj\ during the soft (left) and hard states (right), with different assumptions with respect to the \textit{baseline} SEDs. See text and Ponti et al. (2017) for details.}
\label{moreseds}
\end{figure*}

In order to investigate the effects of our assumptions on the baseline SEDs, we constructed other SEDs as follows. During the soft state the unabsorbed X-ray flux lies within the range $F_{2-10}=4.3-5.7 \times 10^{-10}$ erg cm$^{-2}$ s$^{-1}$ (Ponti et al. 2017). From figure 2 of \citet{Migliari2006}, we derive a radio flux at 8.5 GHz of 0.09 mJy, would the source be in the hard state. We expect that the radio flux might be quenched, in the soft state, therefore we may assume $F_{8.5} = 0.009$ mJy, in any case larger than what assumed in the baseline SED (see `RL' SED in the left panel of Fig.~\ref{moreseds}). Ponti et al. (2017) performed a very detailed analysis of simultaneous \xmm-\nustar\ observations of \axj, adopting different models. We decided to test here two of them, both characterised by a multi-coloured disc and a relativistic iron line, plus a third component, either a blackbody (dubbed `dbb-bb-dl' in Ponti et al, 2017), or a Comptonization model (dubbed `dbb-nth-dl' in Ponti et al, 2017). All the soft state SEDs become undistinguishable, as expected, above $\nu\sim10^{18}$ Hz, being constrained by the X-ray data. 

As for the hard state, the unabsorbed X-ray flux lies within the range $F_{2-10}=4.4-4.9 \times 10^{-11}$ erg cm$^{-2}$ s$^{-1}$ (Ponti et al. 2017), so we derive a radio flux at 8.5 GHz as low as 0.002 mJy, considering the spread and uncertainty on the slope of the relation in \citet{Migliari2006}. This is significantly lower than what assumed in the baseline SED: we refer to this SED as `RQ' in the right panel of Fig.~\ref{moreseds}. Ponti et al. (2017) found no high energy cutoff in the \nustar\ spectra, so that any high energy cutoff, if present, would be located beyond the \nustar\ energy band. We can therefore assume a conservative value of 100 keV for the cutoff in the SED we dub `cutoff' in Fig.~\ref{moreseds}. Finally, Ponti et al. (2017) observed that any inner disc temperature in the range $kT=0.2-0.7$ keV and Comptonisation fractions $f_{sc}=0.3-1.0$ are consistent with the data. We therefore constructed other two extreme SEDs: one with $f_{sc}=1.0$ and $kT=0.7$ keV, and the other with $f_{sc}=0.3$ and $kT=0.4$ keV, dubbed `hot' and `starved', respectively, in Fig.~\ref{moreseds}.

\section{The stability curves}

A photoionized gas will reach an equilibrium at an ionization parameter\footnote{We use here the original \citet{Tarter1969} definition, where $L$ is the total luminosity in ionizing photons, $n$ is the hydrogen density and $r$ is the distance of the gas from the illuminating source.} $\xi=L/nr^2$ and temperature $T$, as a consequence of competing heating and cooling processes depending on its physical and chemical properties, and the illuminating radiation field. It is customary to plot these equilibrium states in a \textit{stability curve} in a $T-\xi/T$ diagram \citep[e.g.][]{Krolik1981}. Assuming $L/r^2$ constant, the latter parameter, $\xi/ T\sim L/r^2 nT \sim 1/p$, is a proxy for the pressure $p$ of the gas. In this diagram, the region on the left-hand side of the stability curve is dominated by cooling processes, while the right-hand side by heating processes. Therefore, an isobaric perturbation (i.e. a vertical displacement in the diagram) of an equilibrium state where the slope of the curve is positive will revert the gas to the same equilibrium state, since an increase in $T$ will move the gas to the cooling region, and the opposite for a decrease in $T$. Such equilibrium states are thus thermally stable. On the other hand, if the slope of the stability curve is negative, any perturbation tends to be amplified (an increase in $T$ will move the gas where heating processes dominate, leading to a further increase of the temperature, and vice versa): the gas is thermally unstable and is likely to collapse rapidly into a different stable equilibrium state.

In the last few years, the stability curves of winds in X-ray binaries have been presented in some papers \citep[e.g.][]{Chakravorty2013a,Chakravorty2016,Higginbottom2015,Higginbottom2016,Dyda2016}. In particular, these studies found that the wind becomes thermally unstable when the illuminating SED is that of a standard hard state, explaining the lack of detections of Fe K absorption in this state. Here we compute stability curves for the gas responsible for the observed absorption features in \axj, adopting the observed SEDs as derived in Sect.~\ref{seds}. We use \textsc{cloudy} 13.03 \citep[last described in][]{Ferland2013} for the computations, adopting Solar abundances (as in Table 7.1 of \textsc{cloudy} documentation), a density $n=10^{12}$ cm$^{-3}$, a turbulence velocity\footnote{The stability curves appear completely insensitive to this parameter in the explored range (from 0 to $2\,000$ km s$^{-1}$), in both states.} $v_{turb}=500$ km s$^{-1}$ and a column density $\log(N_\mathrm{H}/\mathrm{cm}^{-2})=23.5$, as in the photoionization models that best fit the absorption features of \axj\ in \citet{Ponti2015}. As described in the following, we will then investigate the effects on the stability curves of the assumed parameters.

\begin{figure}
\includegraphics[width=\columnwidth]{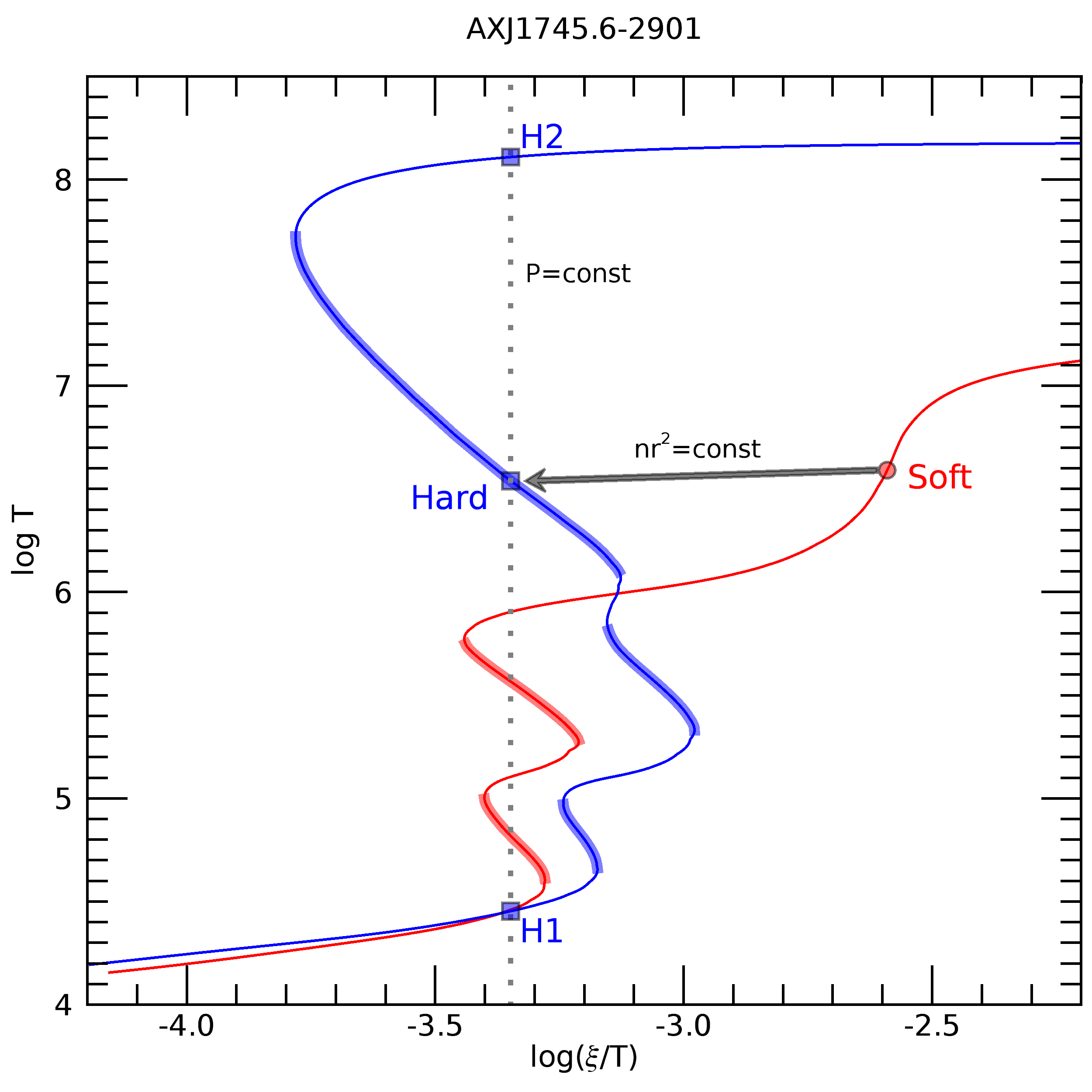}
\caption{Photoionization stability curves once the ionized absorbing plasma observed during the soft state is illuminated with the soft (red line) and the hard (blue line) \textit{baseline} SEDs. The thermally unstable parts of the curves are highlighted with a thicker line. The red circle marks the position of the best-fitting parameters for the observed plasma in the soft state as found by \citet{Ponti2015}, which turns out to be a thermally stable solution. The blue square marks the position of the parameters of the plasma in the hard state, assuming that its physical properties did not change ($nr^2=cost$): the solution is thermally unstable.}
\label{hardsoftcurves}
\end{figure}

\begin{figure}
\includegraphics[width=\columnwidth]{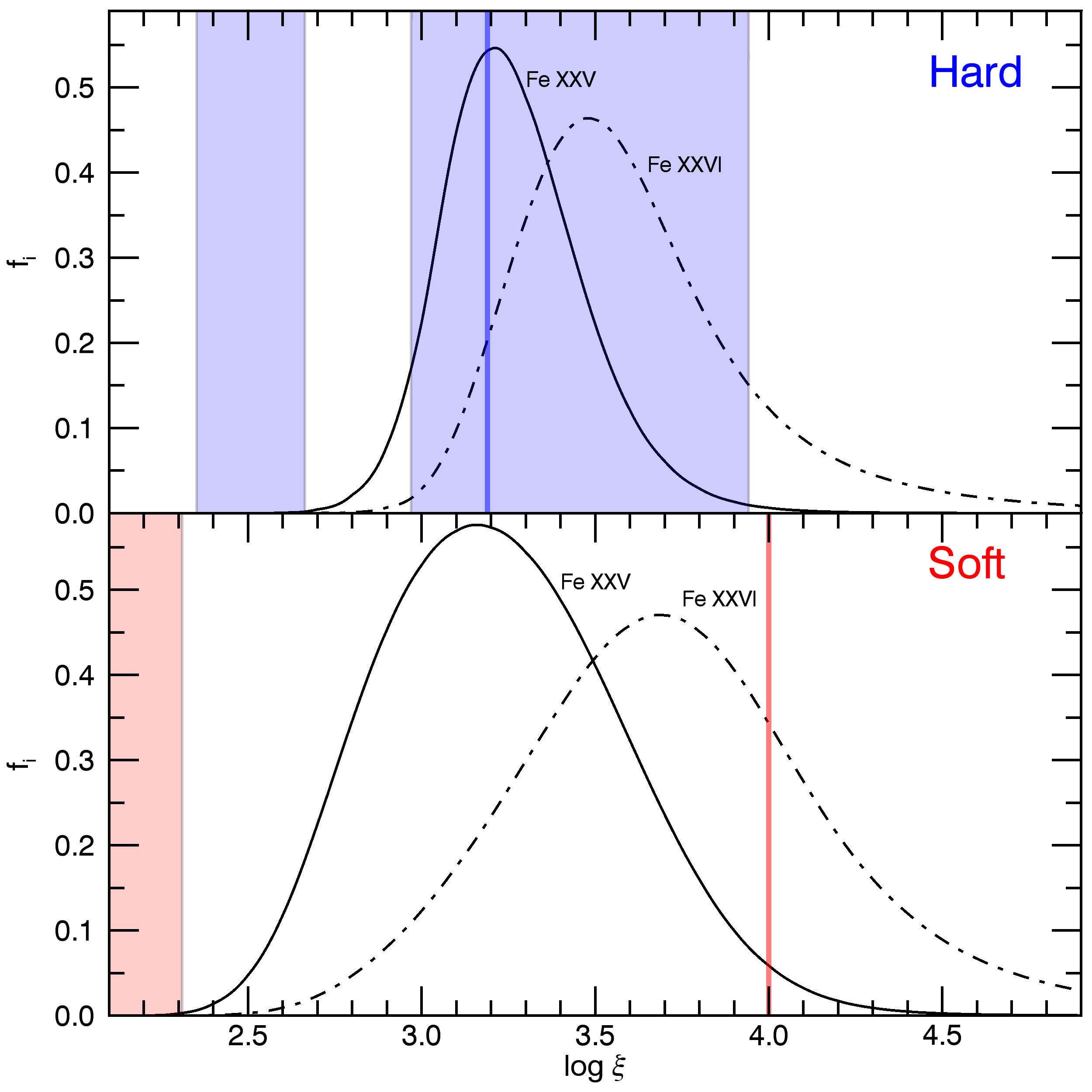}
\caption{Ionic fractions for \Fevc\ and \Fevs, when the illuminating SED is that of the hard (\textit{top}) and soft (\textit{bottom}) state. The red vertical solid line marks the best-fitting value of the ionization parameter for the observed plasma in the soft state as found by \citet{Ponti2015}, while the blue vertical solid line the value of the ionization parameter in the hard state, assuming that its physical properties did not change ($nr^2=cost$). The shaded blue and red bands are the ionization parameter's ranges where the plasma is in the unstable parts of the stability curves shown in Fig.~\ref{hardsoftcurves}.} 
\label{fe_ionization}
\end{figure}

As a first step, we adopted the \textit{baseline} SEDs for the hard and the soft state (see Sect.~\ref{seds} and Fig,~\ref{baselineseds}). The corresponding stability curves are shown in Fig.~\ref{hardsoftcurves}. The plasma observed in the soft state ($\log (\xi_{s}/\mathrm{erg\,s})=4.00$, $T_{s}=3.9\times10^6$ K) clearly lies in a thermally stable branch of the stability curve, as expected. On the other hand, assuming that the physical properties of the plasma did not change in the hard state ($nr^2=const$), the different illuminating SED dramatically changes the stability curve, and the gas would now be in a thermally unstable branch ($\log (\xi_{h}/\mathrm{erg\,s})=3.19$, $T_{h}=3.5\times10^6$ K). Therefore, the gas is very unlikely to be observed in that state, being unstable. On the other hand, the stable branches of the ionization curve for the hard states correspond to ionization parameters characterised by negligible fractions of \Fevc\ and \Fevs. This is clearly shown in Fig.~\ref{fe_ionization}. The plasma observed in the soft state of \axj\ has strong \Fevs\ lines, and is indeed best described by a very high ionic fraction for H-like iron. In any case, all the ionization parameters dominated by \Fevc\ and \Fevs\ are in a stable branch of the stability curve in the soft state. On the other hand, the corresponding values of the ionization parameter are all in unstable branches for the hard state: the absorption features are expected to disappear, as observed.

\begin{figure*}
\includegraphics[width=\columnwidth]{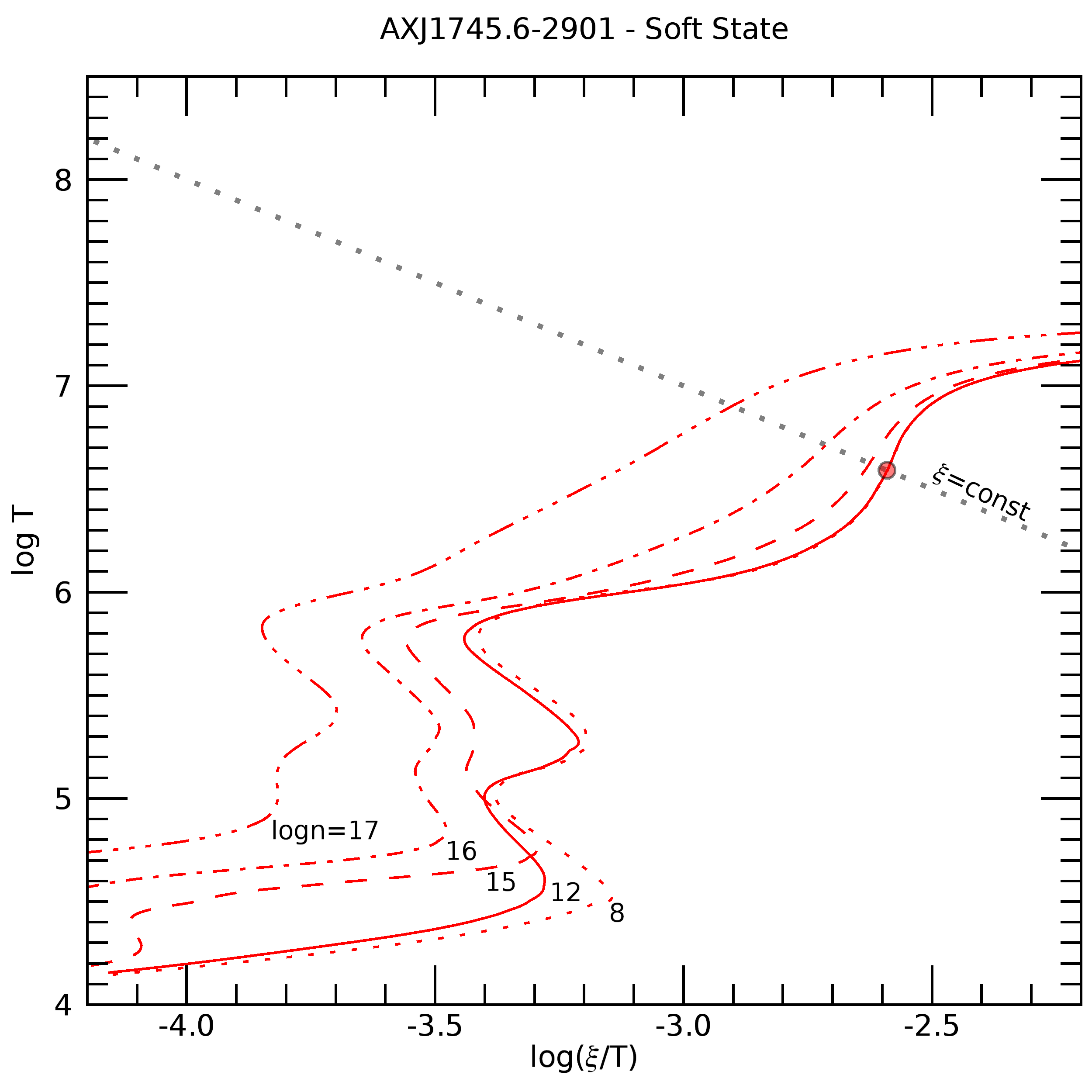}
\includegraphics[width=\columnwidth]{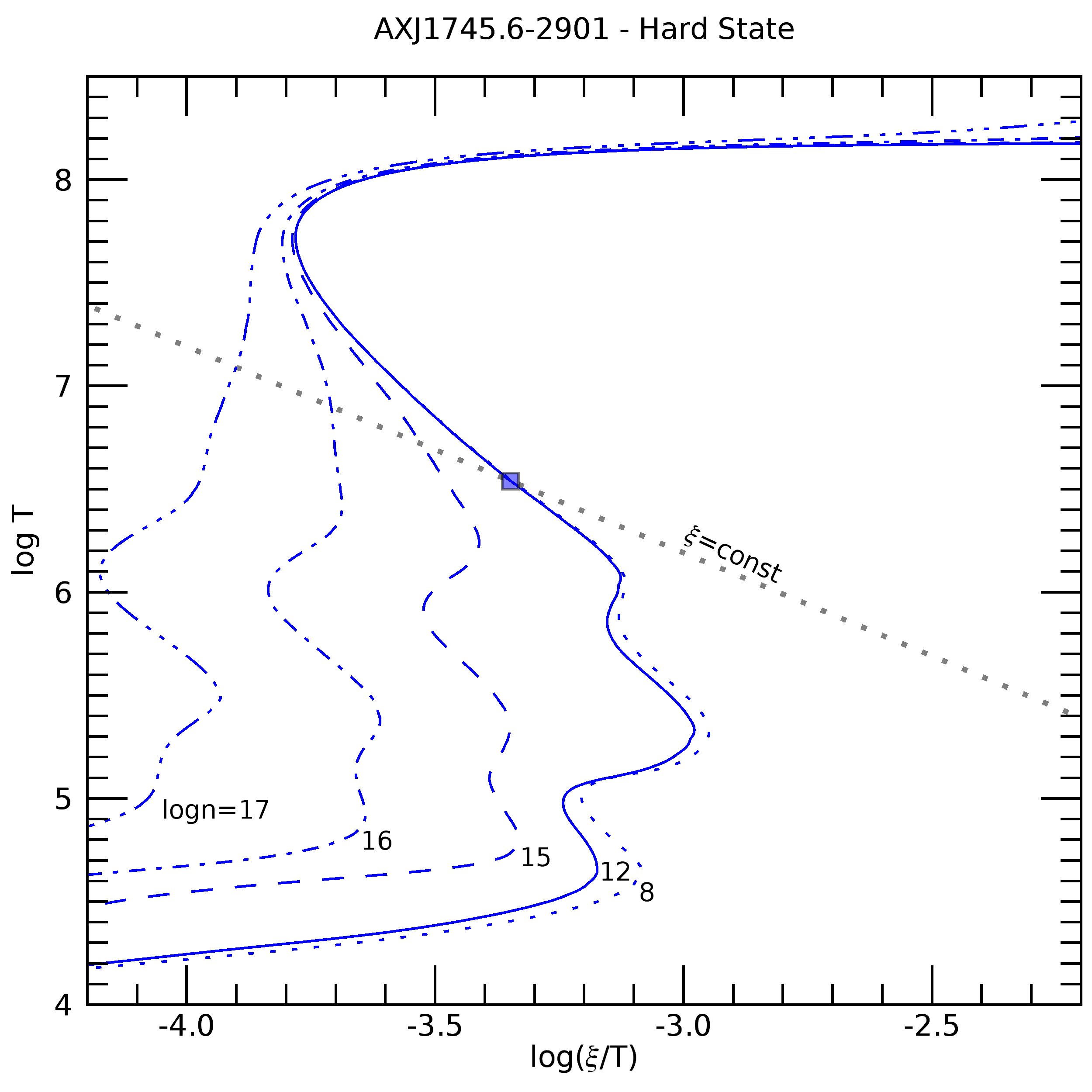}
\caption{Photoionization stability curves once the ionized absorbing plasma is illuminated by the soft (\textit{Left}) and hard (\textit{Right}) state SEDs but for different densities. As in Fig.~\ref{hardsoftcurves}, the red circle marks the position of the best-fitting parameters for the observed plasma in the soft state while the blue square marks the position of the parameters of the plasma in the hard state. The dashed lines in both diagrams have the same value of the ionization parameter $\xi$. The solutions for the soft state are always thermally stable, while those for the hard state are thermally unstable, with the exception of the highest density probed in our simulations.}
\label{dencurves}
\end{figure*}

The photoionization equilibrium of a plasma is in principle uniquely determined by its ionization parameter and the shape of the incident continuum, with no dependence on density. However, the dependence on density of the various competing heating and cooling processes may produce different stability curves at different $n$ for particular SEDs \citep[e.g.][]{Chakravorty2009,Rozanska2008}. We therefore tested the effects of changing the assumed density on the stability curves presented above. The results are shown in Fig.~\ref{dencurves}. In the soft state, the stability curves are very similar for all the densities probed in our simulations. In particular, the plasma parameters are always in a thermally stable solution. On the other hand, for the hard state a change of the slope of the stability curve appears between $10^{17}<n<10^{16}$ cm$^{-3}$: at the highest densities, the branch of the curve where we would expect the plasma to be in the hard state becomes thermally stable. In other words, if the density were at least $10^{17}$ cm$^{-3}$ and all the other plasma parameters were the same as observed in the soft state, the absorption features produced by the plasma should still be visible in the hard state.

\begin{figure*}
\includegraphics[width=\columnwidth]{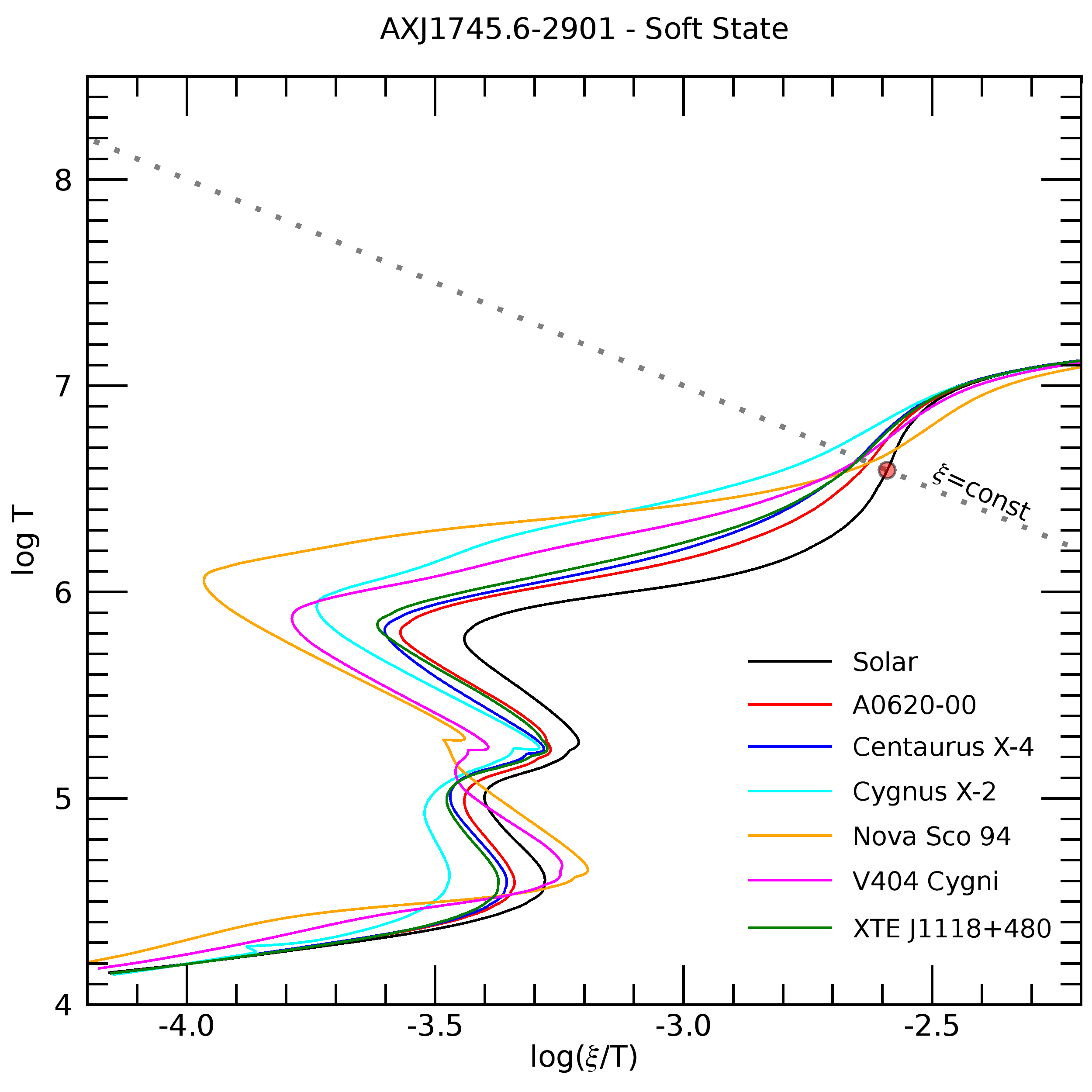}
\includegraphics[width=\columnwidth]{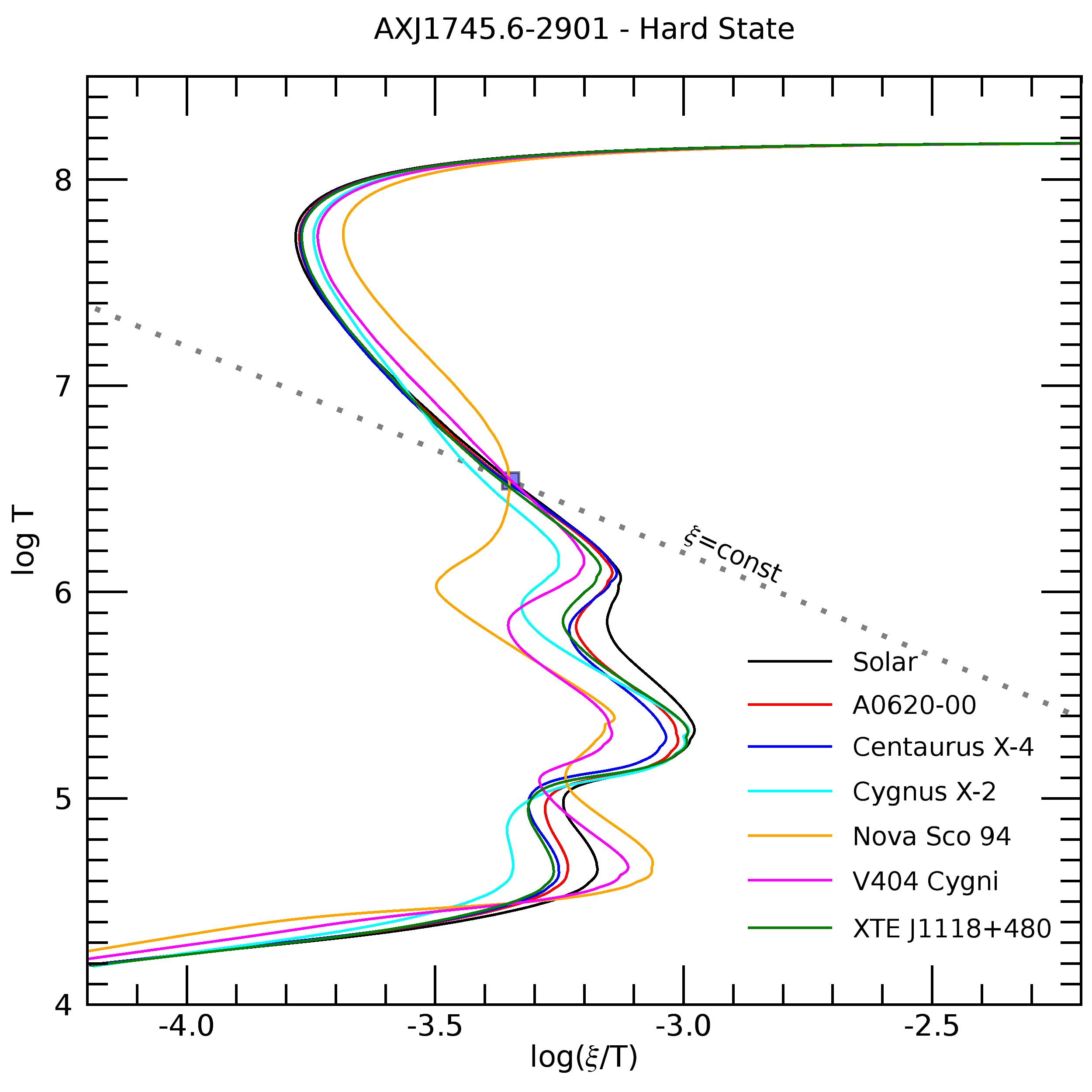}
\caption{Photoionization stability curves when the ionized absorbing plasma has different chemical abundances. As in Fig.~\ref{hardsoftcurves}, the red circle marks the position of the best-fitting parameters for the observed plasma in the soft state while the blue square marks the position of the parameters of the plasma in the hard state. The dashed lines in both diagrams have the same value of the ionization parameter $\xi$. The solutions for the soft state are always thermally stable, while those for the hard state are thermally unstable, with the only exception of the curve relative to the abundance of Nova Scorpii 1994, which is indeed the only one characterised by iron underabundance.}
\label{abundcurves}
\end{figure*}

Different elements constitute important heating/cooling agents at different temperature and ionization parameters. It is therefore expected that the chemical abundances of the gas will yield significant differences in the stability curve. To investigate this dependence, we chose a number of representative chemical abundances for the companion stars in LMXBs \citep[see Table 4 in][]{Casares2016}. The results are shown in Fig.~\ref{abundcurves}. The differences are not very important for the soft state: in particular, the branch of the curve around the observed plasma's properties is not significantly affected by the tested abundances, and always remains stable. This is also true for the curves of the hard state, all being unstable as found above, apart from the curve relative to the abundance of Nova Scorpii 1994 (GRO J1655-40), which is indeed the only one characterised by iron underabundance. In this case, the curve appears just at the end of the stable branch. Therefore, iron underabundance may keep the plasma stable in the hard state, of course at the expense of lower EWs for the iron lines due to the underabundance itself.

\begin{figure*}
\includegraphics[width=\columnwidth]{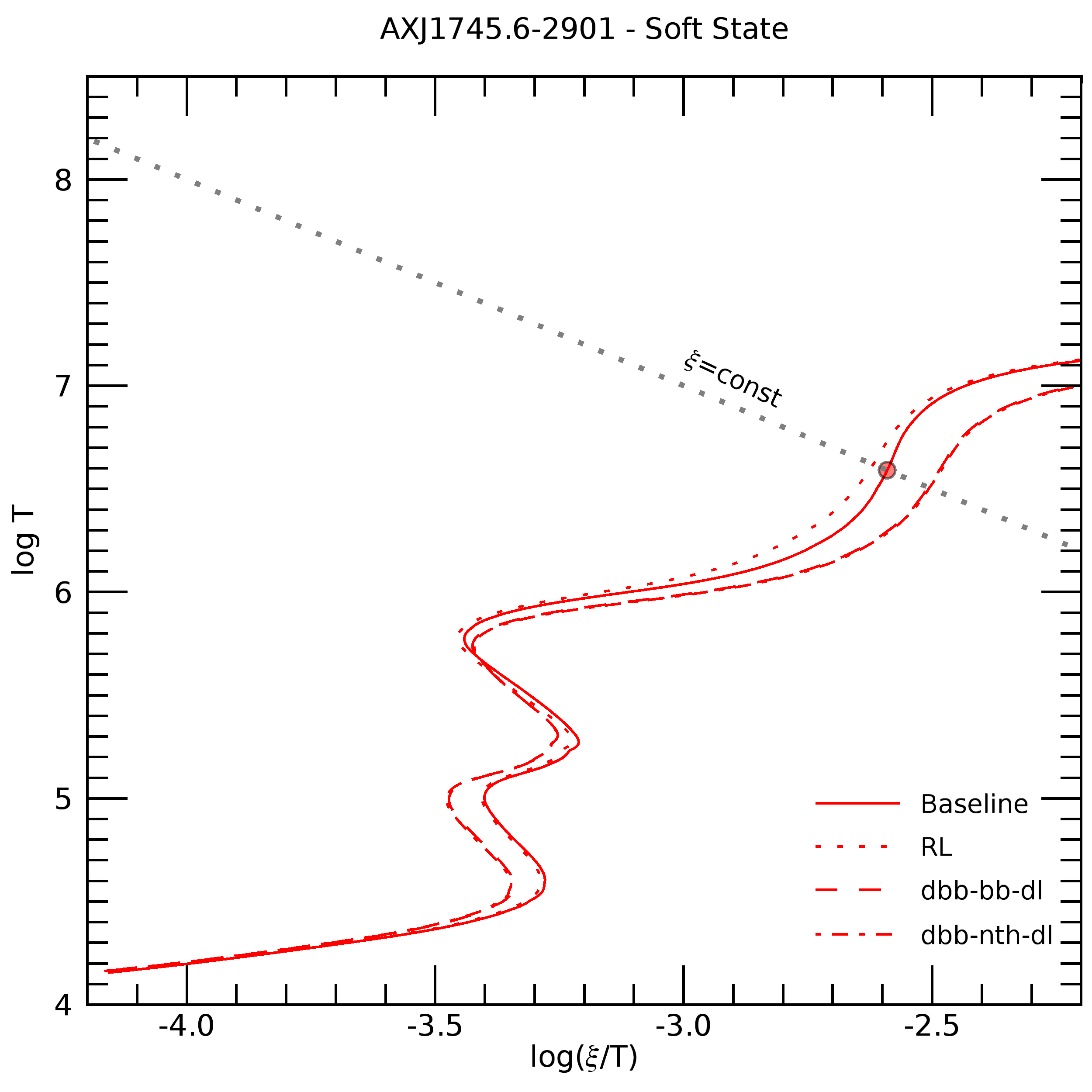}
\includegraphics[width=\columnwidth]{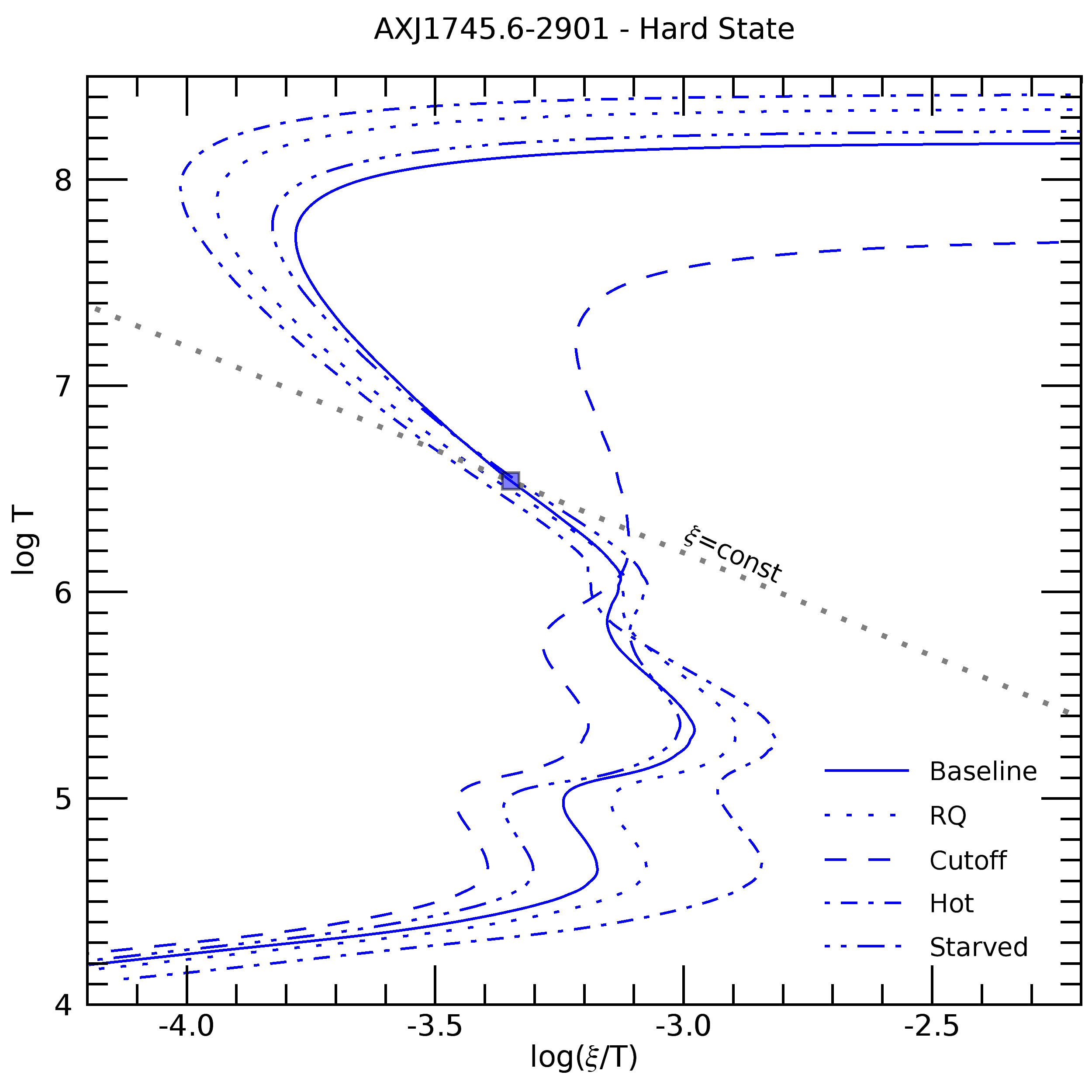}
\caption{Photoionization stability curves when the ionized absorbing plasma is illuminated by the different SEDs described in detail in Sect.~\ref{seds}, both in the soft (\textit{Left}) and in the hard (\textit{Right}) state. As in Fig.~\ref{hardsoftcurves}, the red circle marks the position of the best-fitting parameters for the observed plasma in the soft state while the blue square marks the position of the parameters of the plasma in the hard state. The dashed lines in both diagrams have the same value of the ionization parameter $\xi$. Also in these computations, all the solutions for the soft state are always thermally stable, while those for the hard state are thermally unstable.}
\label{sedcurves}
\end{figure*}

Once we investigated the effects on the stability curves due to the physical parameters of the gas, we further explored the influence on the curve of the different SEDs which can characterize \axj\ both in the hard and in the soft state, as derived in detail in Sect.~\ref{seds}, and shown in Fig.~\ref{moreseds}. The results are shown in Fig.~\ref{sedcurves}. The stability curves for all the SEDs of the soft state are very similar, and the gas is always in a thermally stable solution. In particular, it is interesting to note that, as expected, the inclusion of a radio emission similar to that of the hard SED does not affect the stability of the curve in a significant way. Moreover, the plasma branch remains unstable in the hard state, even if the radio emission is switched off. In other words, the radio emission of the jet is not relevant for the thermal stability of the plasma. On the other hand, the inclusion of a high energy cutoff in the  hard X-rays dramatically changes the stability curve in the parameter region relevant for the iron absorption lines. The lower limit on the cutoff energy compatible with \nustar\ data still keeps thermally unstable that branch of the curve, but it is clear that the curve will tend to the soft state stability curve for lower values of the cutoff. This is a quantitative demonstration that the main difference between the SEDs in the two states affecting the thermal stability of the plasma is represented by the hard X-ray emission.

\section{State transitions}

The results presented in the previous section show that the iron absorption lines observed in \axj\ are compatible with the SEDs observed in the soft state, with no strong dependence on any of the physical and chemical parameters investigated here. On the other hand, they are not compatible with the SEDs of the source in the hard state, due to the enhanced hard X-ray emission. Is this enough to explain the correlation between the state and the presence of the plasma?

In order to take into account only effects related to photoionization thermal stability, let us assume that the absorption lines are produced by a static cloud (with null outflow velocity $v_{out}=0$), placed at a fixed distance $r$ and with fixed extension $\Delta r$, without any external mechanism of creation or replenishment. We can derive a simple estimate of the dimension of the system by using the total mass $M=M_{NS}+M_*\sim2.2$ M$_\odot$, a mass ratio $q=M_{NS}/M_*\sim1.75$, and the orbital period $P=30063.6292$ s \citep{Ponti2017}. From Kepler third law, we get the distance between the NS and the companion star $a\sim9\times10^{11}$ cm, and the Roche lobe can be estimated via the \citet{Eggleton1983} approximation to be $r_L\sim0.43a\sim4\times10^{11}$ cm. Adopting this conservative value as the limiting radial extent of the plasma ($\Delta r \sim r_L$) and assuming a filling factor of unity ($f=1$), the observed column density in the soft state [$\log(N_\mathrm{H}/\mathrm{cm}^{-2})=23.5$] implies a lower limit on the density of the plasma $n>8\times10^{11}$ cm$^{-3}$. In the following, we will therefore assume $n=10^{12}$ cm$^{-3}$, as in our baseline photoionization runs. All the relevant timescales discussed below depend inversely on density in the same way, so they are shorter for larger densities, but their ratios are constant \citep[e.g.][]{Goncalves2007}.

\subsection{Soft to hard}

After the transition from the soft to the hard state, the gas in the plasma will react immediately to the new illuminating SED, since the recombination time at $n=10^{12}$ cm$^{-3}$ is of the order of seconds, and the ionization timescale is even smaller \citep[e.g.][]{Goncalves2007}. Moreover, for a temperature of $T\sim4\times10^6$ K, the thermal timescale is only slightly larger, of the order of tens of seconds \citep[e.g.][]{Goncalves2007}. Therefore, the gas will instantaneously move to the new equilibrium state in the corresponding stability curve, reaching at the same time the ionization and temperature equilibrium, since the two timescales are very similar. It seems then reasonable to assume that, during the transition, both the density $n$ and the distance $r$ of the gas from the illuminating source remain constant. However, as shown in the previous section, this equilibrium state is unstable, and any perturbation will make the gas migrate to a stable solution, on timescales governed by the dynamical timescale:

\begin{equation}
\label{dynt}
t_{dyn} \sim 2\times 10^4 \frac{N_{23}}{T_5^{1/2} n_{12}} \,s
\end{equation}

where $N_{23}$ is the column density in units of $10^{23}$ cm$^{-2}$, $n_{12}$ is the hydrogen density in units of $10^{12}$ cm$^{-3}$, and $T$ is the temperature of the gas in units of $10^5$ K \citep[e.g.][]{Goncalves2007}. Assuming a total column density as in the observed plasma in the soft state ($N_\mathrm{H}=3\times10^{23}$ cm$^{-2}$: Ponti et al. 2017), and the expected temperature $T_{h}=3.5\times10^6$ K of the unstable position just after the transition to the hard state (marked in Fig.~\ref{hardsoftcurves} assuming that $nr^2=const$ during the transition), we get $t_{dyn} \sim1\times10^4$ s. Therefore, starting from this unstable position, in few hours any perturbation will move the gas to a stable solution. Assuming that the average distance $r$ of the gas will not change in this timescale (since $v_{out}=0$), the new stable equilibrium will be characterized by different values of $T$, $\xi$, and $n$.

Considering the stability curve in Fig.~\ref{hardsoftcurves}, in the hard state, several phases of the gas can coexist in pressure equilibrium at values of $\log \xi/T$ around the initial unstable solution. If the plasma moves isobarically from the unstable solution after the transition (i.e a straight vertical displacement, with $\log \xi/T$ kept constant), we have two stable solutions at $\log (\xi_{h1}/\mathrm{erg\,s})=1.10$, $T_{h1}=2.8\times10^4$ K, and $\log (\xi_{h2}/\mathrm{erg\,s})=4.76$, $T_{h2}=1.3\times10^8$ K. Since they are in pressure equilibrium, $n_{h1}T_{h1}=n_{h2}T_{h2}=n_{h}T_{h}$, so $n_{h1}=1.2\times10^{14}$ and $n_{h2}=2.7\times10^{10}$ cm$^{-2}$, if $n_h=10^{12}$ cm$^{-2}$ is the assumed density before and after the transition. There is no easy way to predict which stable solution the plasma will choose: hot and cold clumps can coexist adopting an unknown geometry, or a hot, dilute medium may confine cold, denser clumps, and a part of the cold phase may continuously evaporate to the hot phase and vice versa in a dynamical timescale \citep[e.g.][]{Goncalves2007}.

For simplicity, we may envisage a colder gas condensed in smaller bubbles, with a low filling factor, embedded in a hotter gas with a corresponding larger filling factor. Assuming that the two gas phases occupy the whole available volume (i.e. the filling factors are such that $f_{h1}+f_{h2}=f=1$), the conservation of the total mass leads to:

\begin{align}
\label{h1values}
\begin{split}
f_{h1}=\frac{n-n_{h2}}{n_{h1}-n_{h2}} \sim \frac{n}{n_{h1}} \sim 0.008\\
N_{\mathrm{H}\,h1}=\frac{n_{h1}}{n} f_{h1} N_\mathrm{H} \sim N_\mathrm{H}
\end{split}
\end{align} 

\begin{align}
\label{h2values}
\begin{split}
f_{h2}=\frac{n_{h1}-n}{n_{h1}-n_{h2}} \sim 1\\
N_{\mathrm{H}\,h2}=\frac{n_{h2}}{n} N_\mathrm{H} \sim 0.027 N_\mathrm{H}
\end{split}
\end{align} 

The hot phase has a very high ionization parameter, corresponding to a negligible fraction of \Fevc\ and a very low fraction of \Fevs\ (see upper panel of Fig.~\ref{fe_ionization}), combined with a low column density. Indeed, the result of a computation with \textsc{cloudy}, adopting the hard SED, $\log (\xi_{h2}/\mathrm{erg\,s})=4.76$, $\log (n_{h2}/\mathrm{cm^{-3}})=10.4$ and $\log (N_{\mathrm{H}\,h2}/\mathrm{cm^{-2}})=21.9$ (from Eq.~\ref{h2values}), gives a completely transparent medium, apart from a very weak \Fevs\ K$\alpha$ absorption line with an EW$\sim2$ eV. This is indeed lower than the tightest upper limit found by \citet{Ponti2015} in the hard state. In other words, this component of the plasma will become unobservable.

On the other hand, a computation of \textsc{cloudy} of the cold phase ($\log (\xi_{h1}/\mathrm{erg\,s})=1.10$, $\log (n_{h1}/\mathrm{cm^{-3}})=14.1$) returns a gas which is substantially neutral. The column density expected for this neutral phase should be of the order of the one observed in the soft state, i.e. $\log (N_{\mathrm{H}\,h1}/\mathrm{cm^{-2}})=23.5$ (see Eq.~\ref{h1values}). This is a substantial column density for a neutral gas, capable of absorbing the X-ray emission up to 3 keV. Incidentally, this value is the same as the neutral column density observed in \axj\ both in the hard and in the soft state ($\log (N_\mathrm{H}/\mathrm{cm^{-2}})\simeq23.5$: Ponti et al. 2017). However, the filling factor of this phase is expected to be very low (see Eq.~\ref{h1values}), so it would be only observed as a sporadic change of the persistent neutral absorption. Ponti et al. (2017) report a neutral column density which is constant within few $10^{22}$ cm$^{-2}$ in all the observed hard and soft states. However, dips are excluded in their analysis.

It is well known that the spectra of LMXBs during dipping events are characterised by larger column densities ($\Delta N_\mathrm{H}$ up to $5\times10^{23}$ cm$^{-2}$) and lower ionization parameters ($\Delta \log \xi$ up to -1.5) with respect to persistent states \citep[][and references therein]{DiazTrigo2005}. The spectra during the dips in \axj\ have been indeed fit by \citet{Hyodo2009} with (partial) covering of neutral gas with column densities up to $10^{24}$ cm$^{-2}$. It is therefore difficult, from an observational point of view, to discriminate between the dipping phenomenon and the expected signatures from the cold phase in the hard state.

It is worth noting that the existence of the cold phase depends on the density of the plasma. In Fig.~\ref{dencurves}, the high-temperature branch of the stability curve in the hard state is the same for all densities, so the hot phase is always expected to be present. On the other hand, the low-temperature branch shifts to the left with increasing density. Therefore, the condensation from the unstable solution to the cold phase will move the plasma to the corresponding curve at larger density (as estimated above, $n_{h1}=1.2\times10^{14}$ cm$^{-2}$ if the initial density is $n=10^{12}$ cm$^{-2}$). If the initial density is larger, the stability curve for the cold phase will not be in pressure equilibrium with the hot phase any more, so you may expect only the latter to exist. This will happen when the density of the cold phase approaches $\sim10^{16}$ cm$^{-3}$.

\subsection{Hard to Soft}

At this point we can make the same reasoning as before in the transition from the hard back to the soft state. Assuming again  that $nr^2$ remains constant after the transition, the cold phase will move to $\log (\xi/\mathrm{erg\,s})=1.91$. This is a stable equilibrium for the SED of the soft state (at $T=1.5\times10^5$ K: see Fig.~\ref{hardsoftcurves}). This gas would be only moderately ionized, incapable of impressing iron absorption features to the spectrum (Fig.~\ref{fe_ionization}), but still compatible with the dipping spectra as observed in the soft state. On the other hand, the hot phase will get even more ionized, up to $\log (\xi/\mathrm{erg\,s})=5.57$, a stable solution (although outside the range computed in Fig.~\ref{hardsoftcurves}), but completely transparent. Note that these two phases will not be in pressure equilibrium in the soft state, so, in principle, they will not be confined at the same average distance. Moreover, all the observed soft states always present the same properties ($\xi$, $N_\mathrm{H}$) of the plasma, with small scatter (Ponti et al. 2017). Therefore, even if our simple static model easily explains why the iron absorption lines disappear in the hard state, it does not explain why they re-appear back in the soft state. There must be a mechanism that re-creates the plasma in the soft state, and this mechanism must be self-similar each time, since the plasma properties are always the same in the soft states.

\section{The accretion disc atmosphere?}

The central energies of the iron absorption lines in the \xmm\ spectra of \axj\ are consistent with their rest-frame energies, with upper limits on the outflow velocities of $\sim1\,000$ km s$^{-1}$, basically due to the uncertainties on the energy scale calibration of the EPIC pn (Ponti et al., 2017). 

The observational data are therefore consistent with a static gas in \axj. A possible identification for this gas might be in terms of the atmosphere of the accretion disc.

The observed ionization parameter of the plasma in the soft state ($\log (\xi/\mathrm{erg\,s})=4$) is compatible with a density:

\begin{equation}
n_w\simeq4.8 \,r_{10}^{-2} \times10^{12} \mathrm{cm^{-3}}
\end{equation}

where $r_{10}$ is the distance of the illuminated face of the plasma to the illuminating source (the inner accretion disc), in units of $10^{10}$ cm.
On the other hand, the density of a standard steady $\alpha$ accretion disc can be written as \citep[see e.g.][]{Frank2002}

\begin{equation}
\label{density_ad}
n_{ad}\simeq 6.7 \,r_{10}^{-15/8} \times10^{17} \mathrm{cm^{-3}}
\end{equation}

where we have used a viscosity parameter $\alpha=0.1$, a mass and a radius for the neutron star $M_{NS}=1.4$ M$_\odot$ and $R_{NS}=10$ km, an efficiency $\eta=0.1$ and an Eddington ratio $L/L_{Edd}=0.05$ \citep{Ponti2015}. For $M_{NS}=1.4$ M$_\odot$, $r_{10}$ corresponds to $\sim5\times10^4$ r$_{g}$.

Given the very similar dependence on radius, the ratio between these two densities is approximately constant $n_w/n_{ad}\sim7\times10^{-6}$. Assuming hydrostatic equilibrium, the density of the disc is expected to decrease vertically as \citep[see e.g.][]{Frank2002}

\begin{equation}
n(r,z)=n(r)\exp{(-z^2/2H^2)}
\end{equation}

where $n(r)$ is the density on the central disc plane (as described by \ref{density_ad}), and $H\sim c^2_s R_{NS} \sqrt{R_{NS}/GM_{NS}}$ is the typical scale height of the disc in the $z$-direction, $c^2_s$ being the sound speed. Therefore, a decrease of the density by a factor $n_w/n_{ad}\sim7\times10^{-6}$ corresponds to a height $z\sim4.9\,H$.

If, as proposed above, we identify the plasma in \axj\ as the disc atmosphere, we automatically introduce a mechanism that re-creates it in the soft state, exactly with the same characteristics every time. Indeed, following the transition to the soft state, the accretion disc reverts back to a standard geometrically thin, optically thick disc, and the density of its atmosphere at height $z\sim4.9\,H$ is regulated by the same equations as above. The plasma will therefore acquire the same density and geometrical structure as above, re-producing the same absorption features every time the source is observed in the soft state.

On the other hand, outflow velocities in the range $10^2-10^3$ km s$^{-1}$ have been observed in high inclination BH and NS LMXBs. Indeed, an outflow velocity of just $100$ km s$^{-1}$ would falsify our initial assumption of a static gas, since, in a dynamical timescale the gas could move $\sim10^{11}$ cm, which is comparable to the distance of the gas from the illuminating source. In this case, a launching mechanism for the wind must be invoked in order to re-create the wind in the soft state and replenish it constantly with the same properties. Since the accretion disc itself is expected to change dramatically after the state transitions, it is very likely that this launching mechanism also changes between the soft and the hard state, and this must be taken into account for a global understanding of the plasma behaviour. This approach is attempted in other works \citep[see e.g.][]{Chakravorty2016}, but it is beyond the scopes of this paper, which is only focused on the photoionization properties of the plasma.

\section*{Acknowledgements}

We would like to thank E. Churazov, A. Laor, G.C. Perola, and J. Stern for useful discussions. SB acknowledges financial support from the Italian Space Agency under grant ASI-INAF I/037/12/0, and from the European Union Seventh Framework Programme (FP7/2007-2013) under grant agreement no. 31278. GP acknowledges support from the Bundesministerium f\"ur Wirtschaft und Technologie/Deutsches Zentrum f\"ur Luft- und Raumfahrt (BMWI/DLR, FKZ 50OR 1604) and the Max Planck Society. TMD acknowledges support via a Ram\'on y Cajal Fellowship (RYC-2015-18148).

\bibliographystyle{mnras}
\bibliography{AXJ1745.6-2901}

\begin{thebibliography}{}
\makeatletter
\relax
\def\mn@urlcharsother{\let\do\@makeother \do\$\do\&\do\#\do\^\do\_\do\%\do\~}
\def\mn@doi{\begingroup\mn@urlcharsother \@ifnextchar [ {\mn@doi@}
  {\mn@doi@[]}}
\def\mn@doi@[#1]#2{\def\@tempa{#1}\ifx\@tempa\@empty \href
  {http://dx.doi.org/#2} {doi:#2}\else \href {http://dx.doi.org/#2} {#1}\fi
  \endgroup}
\def\mn@eprint#1#2{\mn@eprint@#1:#2::\@nil}
\def\mn@eprint@arXiv#1{\href {http://arxiv.org/abs/#1} {{\tt arXiv:#1}}}
\def\mn@eprint@dblp#1{\href {http://dblp.uni-trier.de/rec/bibtex/#1.xml}
  {dblp:#1}}
\def\mn@eprint@#1:#2:#3:#4\@nil{\def\@tempa {#1}\def\@tempb {#2}\def\@tempc
  {#3}\ifx \@tempc \@empty \let \@tempc \@tempb \let \@tempb \@tempa \fi \ifx
  \@tempb \@empty \def\@tempb {arXiv}\fi \@ifundefined
  {mn@eprint@\@tempb}{\@tempb:\@tempc}{\expandafter \expandafter \csname
  mn@eprint@\@tempb\endcsname \expandafter{\@tempc}}}

\bibitem[\protect\citeauthoryear{Casares, Jonker  \& Israelian}{Casares
  et~al.}{2016}]{Casares2016}
Casares J.,  Jonker P.~G.,   Israelian G.,  2016, in , Handb. Supernovae.
Springer International Publishing, Cham, pp 1--28 (\mn@eprint {arXiv}
  {1701.07450}), \mn@doi{10.1007/978-3-319-20794-0_111-1}, \url
  {http://link.springer.com/10.1007/978-3-319-20794-0_111-1
  http://arxiv.org/abs/1701.07450}

\bibitem[\protect\citeauthoryear{Chakravorty, Kembhavi, Elvis  \&
  Ferland}{Chakravorty et~al.}{2009}]{Chakravorty2009}
Chakravorty S.,  Kembhavi A.~K.,  Elvis M.,   Ferland G.,  2009, \mn@doi
  [\mnras] {10.1111/j.1365-2966.2008.14249.x}, 393, 83

\bibitem[\protect\citeauthoryear{Chakravorty, Lee  \& Neilsen}{Chakravorty
  et~al.}{2013}]{Chakravorty2013a}
Chakravorty S.,  Lee J.~C.,   Neilsen J.,  2013, \mn@doi [\mnras]
  {10.1093/mnras/stt1593}, 436, 560

\bibitem[\protect\citeauthoryear{Chakravorty et~al.,}{Chakravorty
  et~al.}{2016}]{Chakravorty2016}
Chakravorty S.,  et~al., 2016, \mn@doi [\aap] {10.1051/0004-6361/201527163},
  589, A119

\bibitem[\protect\citeauthoryear{Corbel, Fender, Tzioumis, Nowak, McIntyre,
  Durouchoux  \& Sood}{Corbel et~al.}{2000}]{Corbel2000}
Corbel S.,  Fender R.,  Tzioumis A.,  Nowak M.,  McIntyre V.,  Durouchoux P.,
  Sood R.,  2000, \aap, 359, 251

\bibitem[\protect\citeauthoryear{Corbel, Fender, Tomsick, Tzioumis  \&
  Tingay}{Corbel et~al.}{2004}]{Corbel2004}
Corbel S.,  Fender R.~P.,  Tomsick J.~A.,  Tzioumis A.~K.,   Tingay S.,  2004,
  \mn@doi [\apj] {10.1086/425650}, 617, 1272

\bibitem[\protect\citeauthoryear{Coriat et~al.,}{Coriat
  et~al.}{2011}]{Coriat2011}
Coriat M.,  et~al., 2011, \mn@doi [\mnras] {10.1111/j.1365-2966.2011.18433.x},
  414, 677

\bibitem[\protect\citeauthoryear{{D{\'{i}}az Trigo}, Parmar, Boirin,
  M{\'{e}}ndez  \& Kaastra}{{D{\'{i}}az Trigo} et~al.}{2006}]{DiazTrigo2005}
{D{\'{i}}az Trigo} M.,  Parmar A.~N.,  Boirin L.,  M{\'{e}}ndez M.,   Kaastra
  J.~S.,  2006, \mn@doi [\aap] {10.1051/0004-6361:20053586}, 445, 179

\bibitem[\protect\citeauthoryear{Dyda, Dannen, Waters  \& Proga}{Dyda
  et~al.}{2016}]{Dyda2016}
Dyda S.,  Dannen R.,  Waters T.,   Proga D.,  2016, eprint arXiv:1610.04292

\bibitem[\protect\citeauthoryear{Eggleton}{Eggleton}{1983}]{Eggleton1983}
Eggleton P.~P.,  1983, \mn@doi [\apj] {10.1086/160960}, 268, 368

\bibitem[\protect\citeauthoryear{Fender \& Mu{\~{n}}oz-Darias}{Fender \&
  Mu{\~{n}}oz-Darias}{2016}]{Fender2016a}
Fender R.,  Mu{\~{n}}oz-Darias T.,  2016, in , Vol.~905, Lect. Notes Phys..
Springer International Publishing, pp 65--100 (\mn@eprint {arXiv}
  {1505.03526}), \mn@doi{10.1007/978-3-319-19416-5_3}, \url
  {http://link.springer.com/10.1007/978-3-319-19416-5_3
  http://arxiv.org/abs/1505.03526
  http://dx.doi.org/10.1007/978-3-319-19416-5_3}

\bibitem[\protect\citeauthoryear{Fender, Garrington, McKay, Muxlow, Pooley,
  Spencer, Stirling  \& Waltman}{Fender et~al.}{1999}]{Fender1999}
Fender R.~P.,  Garrington S.~T.,  McKay D.~J.,  Muxlow T. W.~B.,  Pooley G.~G.,
   Spencer R.~E.,  Stirling A.~M.,   Waltman E.~B.,  1999, \mn@doi [\mnras]
  {10.1046/j.1365-8711.1999.02364.x}, 304, 865

\bibitem[\protect\citeauthoryear{Fender, Belloni  \& Gallo}{Fender
  et~al.}{2004}]{Fender2004}
Fender R.~P.,  Belloni T.~M.,   Gallo E.,  2004, \mn@doi [\mnras]
  {10.1111/j.1365-2966.2004.08384.x}, 355, 1105

\bibitem[\protect\citeauthoryear{Fender, Homan  \& Belloni}{Fender
  et~al.}{2009}]{Fender2009}
Fender R.~P.,  Homan J.,   Belloni T.~M.,  2009, \mn@doi [\mnras]
  {10.1111/j.1365-2966.2009.14841.x}, 396, 1370

\bibitem[\protect\citeauthoryear{Ferland et~al.,}{Ferland
  et~al.}{2013}]{Ferland2013}
Ferland G.~J.,  et~al., 2013, Rev. Mex. Astron. y Astrofis., 49, 137

\bibitem[\protect\citeauthoryear{Frank, King  \& Raine}{Frank
  et~al.}{2002}]{Frank2002}
Frank J.,  King A.~R.,   Raine D.~J.,  2002, {Accretion power in
  astrophysics.}.
Cambridge University Press, \url
  {http://adsabs.harvard.edu/abs/2002apa..book.....F}

\bibitem[\protect\citeauthoryear{Gladstone, Done  \&
  Gierli{\'{n}}ski}{Gladstone et~al.}{2007}]{Gladstone2007}
Gladstone J.,  Done C.,   Gierli{\'{n}}ski M.,  2007, \mn@doi [\mnras]
  {10.1111/j.1365-2966.2007.11675.x}, 378, 13

\bibitem[\protect\citeauthoryear{Gon{\c{c}}alves, Collin, Dumont  \&
  Chevallier}{Gon{\c{c}}alves et~al.}{2007}]{Goncalves2007}
Gon{\c{c}}alves A.~C.,  Collin S.,  Dumont A.-M.,   Chevallier L.,  2007,
  \mn@doi [\aap] {10.1051/0004-6361:20066089}, 465, 9

\bibitem[\protect\citeauthoryear{Higginbottom \& Proga}{Higginbottom \&
  Proga}{2015}]{Higginbottom2015}
Higginbottom N.,  Proga D.,  2015, \mn@doi [\apj]
  {10.1088/0004-637X/807/1/107}, 807, 107

\bibitem[\protect\citeauthoryear{Higginbottom, Proga, Knigge  \&
  Long}{Higginbottom et~al.}{2016}]{Higginbottom2016}
Higginbottom N.,  Proga D.,  Knigge C.,   Long K.~S.,  2016, eprint
  arXiv:1612.08996

\bibitem[\protect\citeauthoryear{Hjellming \& Rupen}{Hjellming \&
  Rupen}{1995}]{Hjellming1995}
Hjellming R.~M.,  Rupen M.~P.,  1995, \mn@doi [Nature] {10.1038/375464a0}, 375,
  464

\bibitem[\protect\citeauthoryear{Homan, Neilsen, Allen, Chakrabarty, Fender,
  Fridriksson, Remillard  \& Schulz}{Homan et~al.}{2016}]{Homan2016}
Homan J.,  Neilsen J.,  Allen J.~L.,  Chakrabarty D.,  Fender R.,  Fridriksson
  J.~K.,  Remillard R.~A.,   Schulz N.,  2016, \mn@doi [\apj]
  {10.3847/2041-8205/830/1/L5}, 830, L5

\bibitem[\protect\citeauthoryear{Hyodo, Ueda, Yuasa, Maeda, Makishima  \&
  Koyama}{Hyodo et~al.}{2009}]{Hyodo2009}
Hyodo Y.,  Ueda Y.,  Yuasa T.,  Maeda Y.,  Makishima K.,   Koyama K.,  2009,
  \mn@doi [\pasj] {10.1093/pasj/61.sp1.S99}, 61, S99

\bibitem[\protect\citeauthoryear{Krolik, McKee  \& Tarter}{Krolik
  et~al.}{1981}]{Krolik1981}
Krolik J.~H.,  McKee C.~F.,   Tarter C.~B.,  1981, \mn@doi [\apj]
  {10.1086/159303}, 249, 422

\bibitem[\protect\citeauthoryear{Migliari \& Fender}{Migliari \&
  Fender}{2006}]{Migliari2006}
Migliari S.,  Fender R.~P.,  2006, \mn@doi [\mnras]
  {10.1111/j.1365-2966.2005.09777.x}, 366, 79

\bibitem[\protect\citeauthoryear{Miller et~al.,}{Miller
  et~al.}{2012}]{Miller2012}
Miller J.~M.,  et~al., 2012, \mn@doi [\apj] {10.1088/2041-8205/759/1/L6}, 759,
  L6

\bibitem[\protect\citeauthoryear{Mirabel \& Rodr{\'{i}}guez}{Mirabel \&
  Rodr{\'{i}}guez}{1994}]{Mirabel1994}
Mirabel I.~F.,  Rodr{\'{i}}guez L.~F.,  1994, \mn@doi [Nature]
  {10.1038/371046a0}, 371, 46

\bibitem[\protect\citeauthoryear{Motta, Rouco-Escorial, Kuulkers,
  Mu{\~{n}}oz-Darias  \& Sanna}{Motta et~al.}{2017}]{Motta2017}
Motta S.~E.,  Rouco-Escorial A.,  Kuulkers E.,  Mu{\~{n}}oz-Darias T.,   Sanna
  A.,  2017, \mn@doi [\mnras] {10.1093/mnras/stx570}, 468, 2311

\bibitem[\protect\citeauthoryear{Mu{\~{n}}oz-Darias, Fender, Motta  \&
  Belloni}{Mu{\~{n}}oz-Darias et~al.}{2014}]{Munoz-Darias2014}
Mu{\~{n}}oz-Darias T.,  Fender R.~P.,  Motta S.~E.,   Belloni T.~M.,  2014,
  \mn@doi [\mnras] {10.1093/mnras/stu1334}, 443, 3270

\bibitem[\protect\citeauthoryear{Mu{\~{n}}oz-Darias et~al.,}{Mu{\~{n}}oz-Darias
  et~al.}{2016}]{Munoz-Darias2016}
Mu{\~{n}}oz-Darias T.,  et~al., 2016, \mn@doi [Nature] {10.1038/nature17446},
  534, 75

\bibitem[\protect\citeauthoryear{Mu{\~{n}}oz-Darias et~al.,}{Mu{\~{n}}oz-Darias
  et~al.}{2017}]{Munoz-Darias2017}
Mu{\~{n}}oz-Darias T.,  et~al., 2017, \mn@doi [\mnras] {10.1093/mnrasl/slw222},
  465, L124

\bibitem[\protect\citeauthoryear{Neilsen \& Homan}{Neilsen \&
  Homan}{2012}]{Neilsen2012}
Neilsen J.,  Homan J.,  2012, \mn@doi [\apj] {10.1088/0004-637X/750/1/27}, 750,
  27

\bibitem[\protect\citeauthoryear{Neilsen \& Lee}{Neilsen \&
  Lee}{2009}]{Neilsen2009}
Neilsen J.,  Lee J.~C.,  2009, \mn@doi [Nature] {10.1038/nature07680}, 458, 481

\bibitem[\protect\citeauthoryear{Ponti, Fender, Begelman, Dunn, Neilsen  \&
  Coriat}{Ponti et~al.}{2012}]{Ponti2012b}
Ponti G.,  Fender R.~P.,  Begelman M.~C.,  Dunn R. J.~H.,  Neilsen J.,   Coriat
  M.,  2012, \mn@doi [\mnras] {10.1111/j.1745-3933.2012.01224.x}, 422, L11

\bibitem[\protect\citeauthoryear{Ponti, Mu{\~{n}}oz-Darias  \& Fender}{Ponti
  et~al.}{2014}]{Ponti2014}
Ponti G.,  Mu{\~{n}}oz-Darias T.,   Fender R.~P.,  2014, \mn@doi [\mnras]
  {10.1093/mnras/stu1742}, 444, 1829

\bibitem[\protect\citeauthoryear{Ponti et~al.,}{Ponti et~al.}{2015}]{Ponti2015}
Ponti G.,  et~al., 2015, \mn@doi [\mnras] {10.1093/mnras/stu1853}, 446, 1536

\bibitem[\protect\citeauthoryear{Ponti, Bianchi, Mu{\~{n}}oz-Darias, De, Fender
   \& Merloni}{Ponti et~al.}{2016}]{Ponti2016}
Ponti G.,  Bianchi S.,  Mu{\~{n}}oz-Darias T.,  De K.,  Fender R.,   Merloni
  A.,  2016, \mn@doi [Astron. Nachrichten] {10.1002/asna.201612339}, 337, 512

\bibitem[\protect\citeauthoryear{Ponti, De, Mu{\~{n}}oz-Darias, Stella  \&
  Nandra}{Ponti et~al.}{2017}]{Ponti2017}
Ponti G.,  De K.,  Mu{\~{n}}oz-Darias T.,  Stella L.,   Nandra K.,  2017,
  \mn@doi [\mnras] {10.1093/mnras/stw2317}, 464, 840

\bibitem[\protect\citeauthoryear{R{\'{o}}{\.{z}}a{\'{n}}ska, Kowalska  \&
  Gon{\c{c}}alves}{R{\'{o}}{\.{z}}a{\'{n}}ska et~al.}{2008}]{Rozanska2008}
R{\'{o}}{\.{z}}a{\'{n}}ska A.,  Kowalska I.,   Gon{\c{c}}alves A.~C.,  2008,
  \mn@doi [\aap] {10.1051/0004-6361:200809549}, 487, 895

\bibitem[\protect\citeauthoryear{Russell, Miller-Jones, Maccarone, Yang, Fender
   \& Lewis}{Russell et~al.}{2011}]{Russell2011}
Russell D.~M.,  Miller-Jones J. C.~A.,  Maccarone T.~J.,  Yang Y.~J.,  Fender
  R.~P.,   Lewis F.,  2011, \mn@doi [\apj] {10.1088/2041-8205/739/1/L19}, 739,
  L19

\bibitem[\protect\citeauthoryear{Tarter, Tucker  \& Salpeter}{Tarter
  et~al.}{1969}]{Tarter1969}
Tarter C.~B.,  Tucker W.~H.,   Salpeter E.~E.,  1969, \mn@doi [\apj]
  {10.1086/150026}, 156, 943

\bibitem[\protect\citeauthoryear{Tudose, Fender, Linares, Maitra  \& {Van Der
  Klis}}{Tudose et~al.}{2009}]{Tudose2009}
Tudose V.,  Fender R.~P.,  Linares M.,  Maitra D.,   {Van Der Klis} M.,  2009,
  \mn@doi [\mnras] {10.1111/j.1365-2966.2009.15604.x}, 400, 2111

\makeatother
\end{thebibliography}

\label{lastpage}

\end{document}